\documentclass{aa}   

\usepackage{graphicx}
\usepackage{txfonts}
\usepackage[figuresright]{rotating}
\usepackage{natbib}
\bibpunct{(}{)}{;}{a}{}{,} 

\def\lae{\mathrel{<\kern-1.0em\lower0.9ex\hbox{$\sim$}}}
\def\gae{\mathrel{>\kern-1.0em\lower0.9ex\hbox{$\sim$}}}

\def\mone{$^{-1}$}

\def\rO3HB{$[$OIII$]$5007\slash H$\beta$~}

\def\rN2HA{$[$NII$]$6583\slash H$\alpha$~}
\def\rS2HA{$[$SII$]$\slash H$\alpha$~}

\begin{document}

\title{\ion{H}{i} absorption in \object{3C~49} and \object{3C~268.3}}
\subtitle{Probing the environment of Compact Steep Spectrum and GHz Peaked Spectrum sources}

\author{A. Labiano \inst{1,2}
\and
R. C. Vermeulen \inst{3}
\and
P. D. Barthel\inst{1}
\and
C.P. O'Dea\inst{4}
\and
J. F. Gallimore\inst{5}
\and
S. Baum\inst{6}
\and
W. de Vries\inst{7}
}

\offprints{A. Labiano:  {\tt labiano@astro.rug.nl}}

\institute{Kapteyn Astronomical Institute, Groningen, 9700 AV, The Netherlands 
\and
Space Telescope Science Institute, Baltimore, MD 21218, USA 
\and
Netherlands Foundation for Research in Astronomy, Dwingeloo, The Netherlands
\and
Department of Physics, Rochester Institute of Technology, Rochester, NY, 14623, USA
\and
Department of Physics, Bucknell University, Lewisburg, PA 17837,USA  
\and
Center for Imaging Science, Rochester Institute of Technology,  Rochester, NY 14623. USA 
\and
Lawrence Livermore National Lab., Livermore CA, 94550, USA
}

\date{Received <date> / accepted <date>}

\titlerunning{\ion{H}{i} absorption in 3C~49 and \object{3C~268.3}}
\authorrunning{Labiano et al.}

\abstract{
We present and discuss European VLBI Network UHF band spectral line observations, made to localise the redshifted 21cm HI absorption known to occur in the subgalactic sized compact steep spectrum galaxies \object{3C~49} and \object{3C~268.3}. We have detected \ion{H}{i} absorption towards the western radio lobe of \object{3C~49} and the northern lobe of \object{3C~268.3}. However, we cannot rule out the presence of similar amounts of \ion{H}{i} towards the opposite and much fainter lobes. The radio lobes with detected \ion{H}{i} absorption (1) are brighter and closer to the core than the opposite lobes; (2) are depolarized; and (3) are associated with optical emission line gas. The association between the \ion{H}{i} absorption and the emission line gas, supports the hypothesis that the \ion{H}{i} absorption is produced in the atomic cores of the emission line clouds. Our results are consistent with a picture in which compact steep spectrum sources interact with clouds of dense gas as they propagate through their host galaxy. We suggest that the asymmetries in the radio and optical emission can be due to interaction of a two sided radio source with an asymmetric distribution of dense clouds in their environment.\\


}

\maketitle


\section{Introduction}

The GHz Peaked Spectrum (GPS) and Compact Steep Spectrum (CSS) radio sources make up significant fractions of the extragalactic bright (cm wavelength selected) radio source population ($\sim$10\% and $\sim$20\%, respectively), but are not well understood \citep[e.g.][]{O'Dea98}. They are powerful but compact radio sources whose spectra are generally simple and convex with peaks near 1 GHz and 100 MHz respectively. The GPS sources are entirely contained within the extent of the nuclear narrow line region ($\lae 1$ kpc, NLR) while the CSS sources are contained entirely within the host galaxy ($\lae 15$ kpc). GPS and CSS sources are important because (1) they probe the NLR and interstellar medium (ISM) of the host galaxy and (2) they may be the younger stages of powerful large-scale radio sources -- giving us insight into radio source genesis and evolution.\\

The currently favored hypothesis is that the GPS and CSS sources are indeed the young progenitors of the large scale powerful double sources \citep[e.g.][]{Carvalho85, Hodges87, Begelman96, Fanti95, Readhead96b, O'Dea98, Snellen00, Alexander00}. In this model they propagate relatively quickly through the ISM of the parent galaxy with advance speeds of a few percent of the speed of light. Observed proper motions tend to be a bit higher - in the range 0.05 - 0.2 h\mone c , e.g., \citet{Polatidis03}, though the detections may be biased towards objects with the highest velocities.\\

Searches for 21 cm \ion{H}{i} absorption have produced a 50\% detection rate in GPS and CSS sources \citep{Vermeulen03, Pihlstrom03} in contrast to normal elliptical radio galaxies, where the detection rate does not exceed $\sim$10\% \citep{Gorkom89}. This indicates that clouds of atomic hydrogen are very common in the environments of GPS and CSS sources or that the geometry and/or morphology of GPS and CSS sources is very favorable for the detection of \ion{H}{i} absorption. The close alignment and similar spatial extents of the radio continuum and optical emission line plasma suggests the existence of a close coupling between the thermal gas and the radio sources \citep{Vries97, Vries99, Axon00}. The broad and highly structured spatially integrated [OIII]$\lambda 5007$ line widths observed by \citet{Gelderman94} strongly suggest that the radio source is dominating the emission line kinematics. This has been confirmed by Hubble Space Telescope spectroscopy \citep{O'Dea02}. \\

Thus, the simple picture of evolution may require the incorporation of interaction of the radio sources with dense gas clouds in their ISM (e.g., \citet{Jeyakumar05}). In order to probe the nature of the relationship between the gas clouds and the radio source we have obtained high resolution EVN observations of the redshifted 21 cm line seen in absorption against 1 GPS and 2 CSS radio galaxies in our WSRT spectra \citep{Vermeulen03}. The results for 2050+364 are presented by \citet{Vermeulen05}. Here we present the results for \object{3C~49} and \object{3C~268.3}.\\

\section{Observations and data reduction}
\label{sec:obs}

The 21cm \ion{H}{i} absorption lines associated with the sources \object{3C~49} ($z=0.6207$, 876.7 MHz) and \object{3C~268.3} ($z=0.37116$, 1035.1 MHz)\footnote{Redshifts from \citet{Spinrad85} for \object{3C~49} and \citet{Gelderman94} for \object{3C~268.3}.} were observed on 1999 September 14 and 09, respectively, for about 7 hours each, using the UHF receivers (800--1300 MHz) on the European VLBI Network (EVN). The sources \object{J0249+063}, \object{3C~84}, \object{J1048+717}, \object{3C~286}, \object{DA~406}, and \object{3C~454.3} were used as calibrators. The recorded bandwidth of 4 MHz was correlated at the NRAO, Socorro correlator with 256 spectral channels, for a resolution of 5.3 km\,s$^{-1}$ and 4.5 km\,s$^{-1}$ in \object{3C~49} and \object{3C~268.3}, respectively. \\

At the time of these observations, we could only obtain data in left and right circular polarisations from the Effelsberg and Westerbork (Tied Array) telescopes, and two orthogonal linear polarisations from the Onsala (25m) telescope. For \object{3C~268.3}, we obtained left circularly polarised data from the Lovell Telescope at Jodrell Bank in addition. \\

The NRAO Astronomical Image Processing Software (AIPS) package was used for the initial data processing (fringe fitting, spectral passband calibration and a priori gain calibration). The Caltech DIFMAP package \citep{Shepherd97} was used for all further calibration and analysis (Cleaning, self-calibration, model-fitting). The data processing and analysis steps used to obtain final continuum images and radio spectra towards the various features in these images are explained in detail in our 2050+364  EVN UHF observations paper \citep{Vermeulen05}. It should be noted in particular that all available polarization products were averaged together, including the cross-correlations between linear and circular polarisations. We believe the sensitivity gained to total intensity is more important than resultant limitations on absolute flux calibration accuracy or on (image or spectral) dynamic range; we think these are more affected by the sparseness of the array and the lack of complete system temperature data and gain curves. In fact, we caution that the absolute flux density scales for the data shown are uncertain even at the 50\% level\footnote{This error is not included in our reported errors on flux densities.}. However, the main astrophysical results are unaffected, since they depend on opacities rather than on absolute flux densities. Furthermore, great effort was expended to obtain reliable {\it relative} (self)calibrations between the telescopes, in an extensive series of very gradual self-calibration iterations. \\

\begin{table}
\begin{minipage}{\columnwidth}
\caption{Observational parameters of the sources.}
\label{obstab}
\centering
\resizebox{\hsize}{!}{
\begin{tabular}{cccccccc}

\hline
\hline

  & Freq. & Time on source  & Number & RMS & Cont. Peak & \multicolumn{2}{c}{Restoring beam}\\ 
\cline{7-8}
Source & (GHz). & hh:mm:ss & of scans &(Jy beam\mone) & (Jy beam\mone) & (mas) & (deg) \\

\hline

\object{3C~49}    &0.8767 &  07:26:18 & 4 & $2.5\times10^{-3}$ & 6.550 & 105 x 51.5  & -45.6 \\
\object{3C~268.3} & 1.0351 & 06:59:14 & 4 & $1.4\times10^{-3}$ & 0.601 & 70.2 x 43.5 & 59.7 \\
\hline
\end{tabular}
}
\end{minipage}
\\  \\
Summary of the EVN observations of \object{3C~49} and \object{3C~268.3}, in September, 1999. The first column gives the name of the source, the second column the frequency of our observations, the third column the exposure time, the fourth the number of scans made on each particular source (the exposure time shown is the sum of all the scans). The RMS column is the noise in an empty region of the map. {\it Cont. Peak} is the peak flux density of the continuum image. The {\it Beam} column gives the FWHM and orientation of the beam.

\end{table}

\begin{figure}
\centering
\includegraphics[width=\columnwidth]{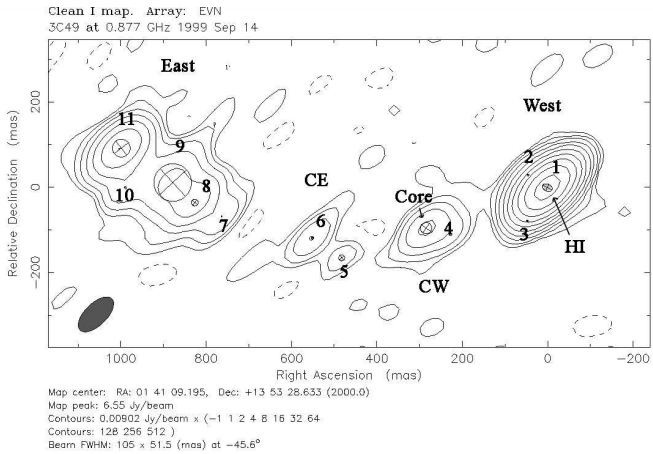}
\caption{Clean map of \object{3C~49} with the numbered Gaussian components drawn on top. We use letters for 
the major source regions (from left to right: E, CE, CW, W). The core of the source is in the center of the {\it CW} component \citep{Ludke98}.\label{3c49mod}} 
\end{figure}

\begin{figure}
\centering
\includegraphics[width=\columnwidth]{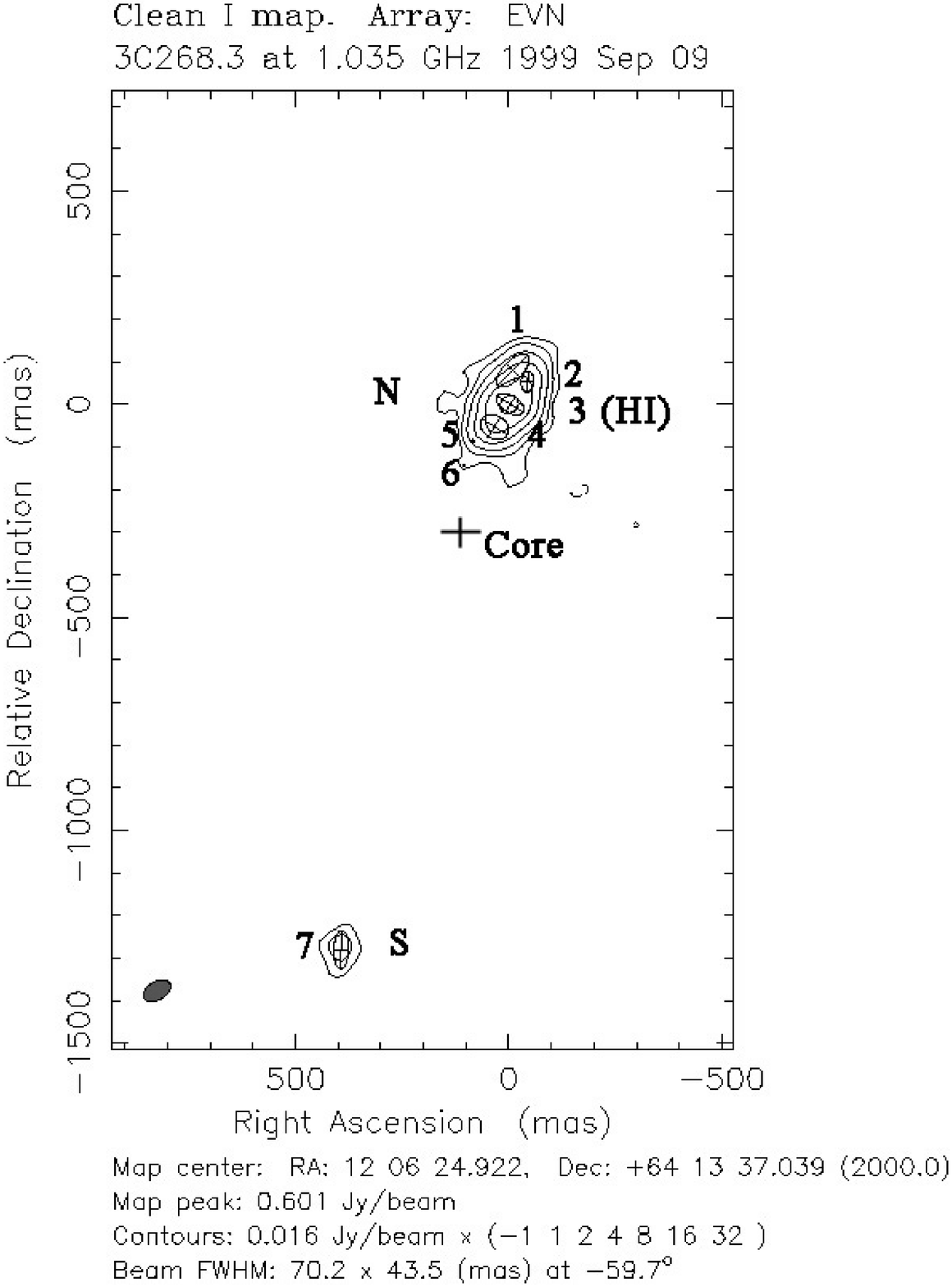}
\caption{ Clean map of \object{3C268.3} with the Gaussian components drawn on top. We use numbers to refer to the Gaussian model components (1 to 7, from top to bottom, being 1 the furthest north, and 7 the furthest south) and letters for the source {\it real} components (N for the northern, S for the southern ). The position of the core detected at 5 GHz by \citep{Ludke98} has been marked with a cross.\label{3c268mod}} 
\end{figure}

We believe the continuum structures and spectral line profiles obtained are robust against the overall calibration uncertainties. Indeed, the continuum structures obtained at these novel frequencies correspond well with those found at other frequencies, as discussed in Section~\ref{subsec:continuum}.  In order to restrict the number of free parameters, the sky model fitted to the visibility data during the self-calibration cycles consisted of a limited number of Gaussian components, which are shown overplotted on the continuum images in Figures \ref{3c49mod} and \ref{3c268mod}; their parameters, fitted to the visibility data, are given in Tables~\ref{measurements1} and \ref{measurements2}. \\

\begin{table*}
\begin{minipage}{\columnwidth}
\caption{Components and absorption line properties. Detections.}
\label{measurements1}
\centering
\resizebox{2\hsize}{!}{
\begin{tabular}{ccccccccccccc}
\hline
\hline
Source & GC & R & $\theta $ & Major& Minor & Flux$^a$ &  SC & $\Delta $S/S$_{cont}$ & FWHM & $\tau$ $^b$ & $N_\mathrm{\ion{H}{i}}$ $^b$ & Center \\
 &  & (mas) & (deg) & (mas) & (mas) & (Jy) &   &  & (km/s) &  & 10$^{20}$ ($T_\mathrm{S}$/100K) cm$^{-2}$ & (km/s) \\
\hline
\object{3C~49}    & 1 (W) & 1.24 & 133 & 24&24& 7.28$\pm$0.03 & 1 &    3.5\% $\pm$ 0.4\% & 20.9$\pm$3.7 & 0.036$\pm$0.003& 1.5$\pm$0.3 &-138 $\pm$2 $\pm$19\\
\object{3C~49}    & 1 (W) &  -      &  -       &  - & - &                -              & 2 &    1.9\% $\pm$ 0.4\% & 22.7$\pm$8.1& 0.019$\pm$0.006& 0.8$\pm$0.4 & -160 $\pm$5 $\pm$19\\
\object{3C~268.3}& 3 (N) & 3.9 & -88 &72&39& 1.10$\pm$0.01 & - & 2.5\% $\pm$ 0.6\% & 67.1$\pm$6.0 & 0.025$\pm$0.005 & 3.2$\pm$0.6 & +190 $\pm$2 $\pm$12\\
\hline 
\end{tabular}
}
\end{minipage}
\\ \\
Columns 2 to 7 list the properties of the fitted Gaussian components ({\it GC}): distance (to coordinates (0,0)) and position ({\it R, $\theta$}), axes' size ({\it Major, Minor}), and integrated flux density. Columns 8 to 13 list the properties of the \ion{H}{i} absorptions: spectral component ({\it SC}) : peak depth divided by continuum level ({\it $\Delta$S/S$_{cont}$}), width ({\it FWHM}), optical depth ($\tau$), column density ({\it $N_\mathrm{\ion{H}{i}}$}) and central wavelength ({\it Center}).\\
$^a$: The errors in the absolute flux density scales include only the formal statistical error, and do not include the errors due to sparseness of the array (up to 50\%, see Section \ref{sec:obs}).\\
$^b$: Not taking into account the covering factor (See section \ref{sec:absorption}).\\
\end{table*}



\begin{table*}
\begin{minipage}{\columnwidth}
\caption{Components and absorption line properties. Non detections}
\label{measurements2}
\centering
\resizebox{2\hsize}{!}{
\begin{tabular}{ccccccccccc}

\hline
\hline
Source & MC & R & $\theta $ &Major&Minor& Flux$^a$ &  $\Delta $S/S$_{cont}$ & FWHM & $\tau$  $^b$& $N_\mathrm{\ion{H}{i}}$  $^b$ \\
 &  & (mas) & (deg) & (mas) & (mas) & (Jy) &    & (km/s) &  & 10$^{20}$ ($T_\mathrm{S}$/100K) cm$^{-2}$ \\
\hline

\object{3C~49}  & 2 (W)    & 54.5 &  58  &4&4& 0.13$\pm$0.02 &  $  <31\%$ & 21.8 & $<0.37$ & $<16$  \\
\object{3C~49}  & 3 (W)    & 92.6 & 149 &4&4& 0.05$\pm$0.02 & $<100\%$ & 21.8 &       X       &    X        \\
\object{3C~49}  & 4 (CW) & 300  & 108 &40&40& 0.23$\pm$0.02 &   $<19\%$ & 21.8 & $<0.21$ & $<8.7$  \\
\object{3C~49}  & 5 (CE)  & 509  & 109 &14&14& 0.04$\pm$0.02 & $<100\%$ & 21.8 &       X       &      X      \\
\object{3C~49}  & 6 (CE)  & 564  & 102 &6&6& 0.07$\pm$0.02  &   $<57\%$ & 21.8 & $<0.85$ & $<36$  \\ 
\object{3C~49}  & 7 (E)     & 767  &   95 &3&3& 0.07$\pm$0.02  &   $<57\%$ & 21.8 & $<0.85$ & $<36$  \\
\object{3C~49}  & 8 (E)     & 827  &   93 &16&16& 0.21$\pm$0.03 &   $<30\%$ & 21.8 & $<0.36$ & $<15$  \\
\object{3C~49}  & 9 (E)     & 878  &   89 &88&88& 0.46$\pm$0.04  &   $<17\%$ & 21.8 & $<0.19$ & $<8.1$  \\
\object{3C~49}  & 10 (E)   & 989  &   90 &5&5& 0.11$\pm$0.02  &   $<36\%$ & 21.8 & $<0.45$ & $<19$  \\
\object{3C~49}  & 11 (E)   &1003&   85  &41&41& 0.81$\pm$0.03 &     $<7.4\%$ & 21.8 & $<0.08$ & $<3.3$  \\
\object{3C~49} & East Int. &   --   &    --   &--&--& 1.66$\pm$0.04 &    $<5.0\%$  & 21.8 &  $<0.05$ & $<2.1$ \\
\object{3C~49} & CE Int.   &   --   &    --   &--&--& 0.11$\pm$0.04  &    $<63\%$  & 21.8 &  $<1.00$ & $<42$ \\

\object{3C~268.3} & 1 (N) &   80 &  -7.2 &100&45& 0.24$\pm$0.01 &$<10\%$&  67.1 &$<0.11$&$<14$\\
\object{3C~268.3} & 2 (N) &   69 &  -40  &49&30& 0.31$\pm$0.01  &$<6.0\%$&    67.1 &$<0.06$&$<8.1$\\
\object{3C~268.3} & 4 (N) &   62 &  148 &71&55& 0.57$\pm$0.02 &$<3.0\%$&    67.1 &$<0.03$&$<4.0$\\
\object{3C~268.3} & 5 (N) & 177 &  143 &3&3& 0.020$\pm$0.007 &$<56\%$&  67.1 &$<0.82$&$<107$ \\
\object{3C~268.3} & 6 (N) & 120 &  135 &6&6& 0.01$\pm$0.01 &$<100\%$&67.1 &      X      &     X     \\
\object{3C~268.3} & 7 (S) & 1339& 162 &90&35& 0.104$\pm$0.007 &$<13\%$&  67.1 &$<0.14$&$<18$\\

\hline
\end{tabular}
}
\end{minipage}
\\ \\
As for Table \ref{measurements1} but for non detections. The numbers shown correspond to the 2$\sigma$ detection limits (maximum peak depth = 2$\times$RMS/continuum). An "X" represents no value, as the absorption could be $100\%$. Rows 11 and 12, {\it East Int, CE Int}, correspond to the spectrum spatially integrated over the entire Eastern and CE component of \object{3C~49} respectively. For \object{3C~268.3}, the South component consists of a single Gaussian component (7).\\
$^a$: The errors in the absolute flux density scales include only the formal statistical error, but do not include the errors due to sparseness of the array (up to 50\%, see Section \ref{sec:obs}).\\
$^b$: Not taking into account the covering factor (see Section \ref{sec:absorption}).\\

\end{table*}

\subsection{Spectrum analysis}

\begin{figure}
\centering
\includegraphics[width=\columnwidth]{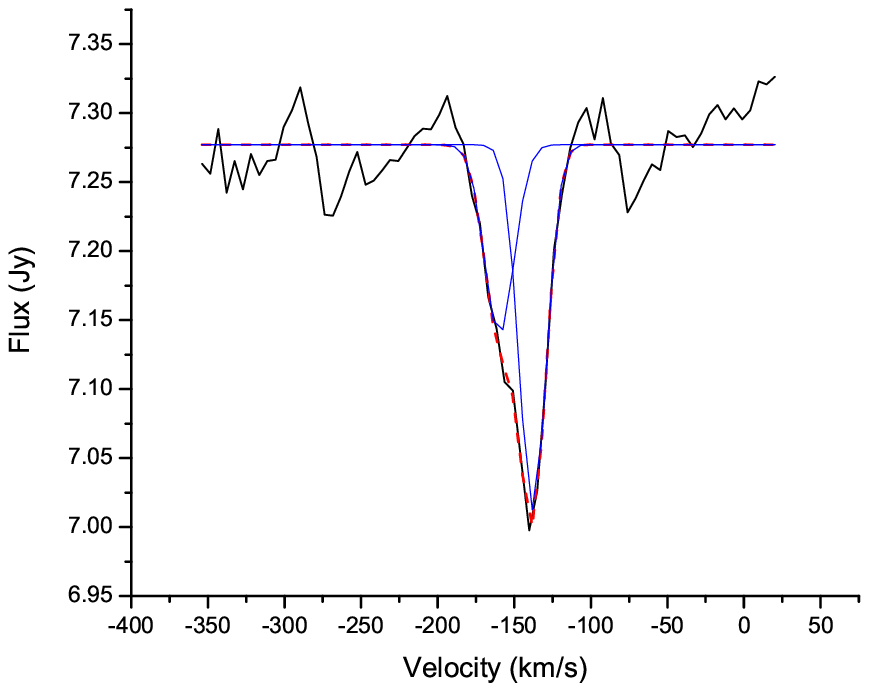}
\caption{Spectrum of the absorption in the West component of \object{3C~49}, with the de-blended lines and total fit over plotted. \label{spec1}}
\end{figure}

\begin{figure}
\centering
\includegraphics[width=\columnwidth]{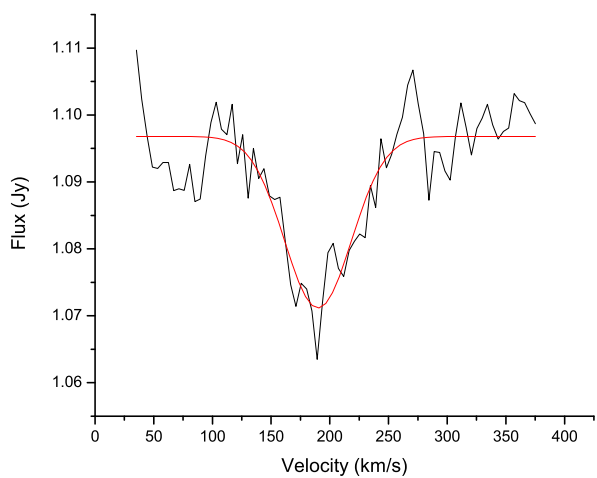}
\caption{Spectrum of the absorption in the North component of \object{3C~268.3}, with the fit over-plotted. \label{spec3}}
\end{figure}

\begin{figure}
\centering
\includegraphics[width=\columnwidth]{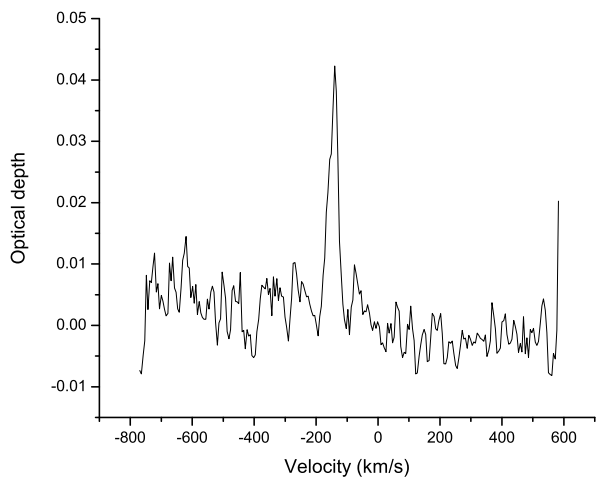}
\caption{Optical depth for the for the \ion{H}{i} detection in the West component of \object{3C~49}. \label{tau1}}
\end{figure}

\begin{figure}
\includegraphics[width=\columnwidth]{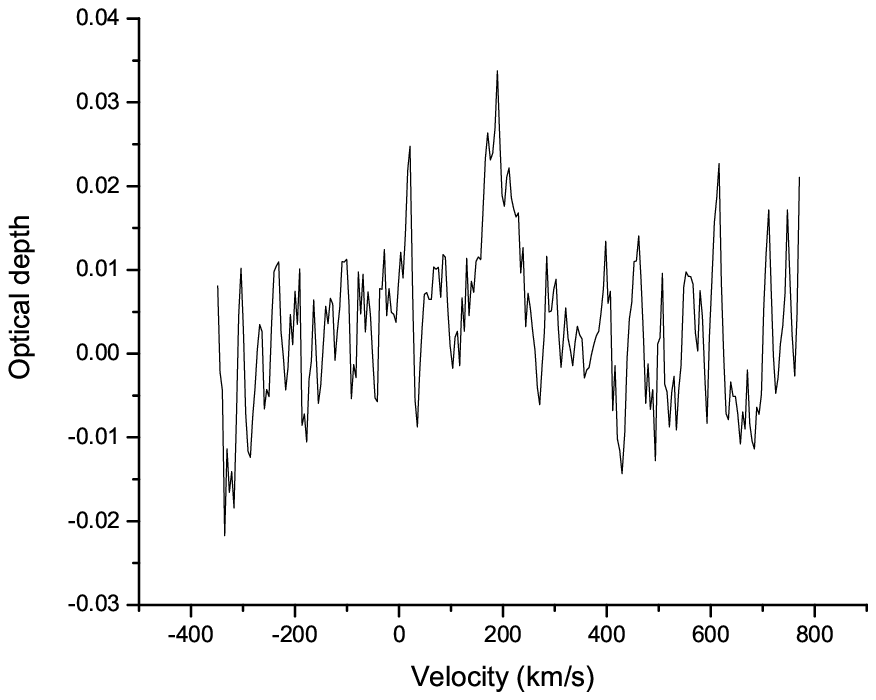}
\caption{Optical depth for the \ion{H}{i} detection in the North component of \object{3C~268.3}.\label{tau3}}
\end{figure}

Given the sparseness of these VLBI datasets, the most robust way to derive line spectra of the various regions of the sources was to re-fit, separately for each of the line channels, the flux density of each of the Gaussian model components, as first derived for the line-free continuum. The positions and diameters of the model components were fixed at the continuum values. Resultant spectra are shown in Figures \ref{spec1} to \ref{tau3} for the locations where absorption was detected. Showing percentage absorption depth or opacity with respect to the continuum strength of the appropriate components avoids the uncertainties in the absolute flux density scale discussed above. Zero velocity was taken to be at the nominal optical redshift found in the literature; all calculations used the appropriate relativistic formulae. \\

%
%
%
%

We fitted Gaussian profiles to the absorption towards our sources using standard, non-linear $\chi^2$ minimization techniques. We measured the optical depth of each line (Figures \ref{tau1} and \ref{tau3}), its full width at half maximum (FWHM), and its centroid. The values are listed with their formal fitting errors in Tables \ref{measurements1} and \ref{measurements2}. The reduced $\chi^2$ indicated that two Gaussian lines were needed to fit the absorption towards \object{3C~49} (Figure \ref{spec1}), while one was sufficient for \object{3C~268.3} (Figure \ref{spec3}). For all locations where the spectra showed no detectable absorption, 2$\sigma$ limits have been estimated from the spectral r.m.s.\ noise level, assuming that any putative line would occur at a similar velocity as the line detected for the same source, and with the same FWHM as for the detection (the average of the two detections for \object{3C~49}). The resultant limits are given in Table \ref{measurements2}. \\

\section{Results}
\label{sec:results}

We have detected localized 21 cm \ion{H}{i} absorption regions in both \object{3C~49} and \object{3C~268.3} consistent with the previously published WSRT observations \citep{Vermeulen03}. The higher resolution of our VLBI observations allows us to locate the \ion{H}{i} absorber and propose a physical scenario. We have studied the continuum milliarcsecond structure of the sources, as well as the properties of the \ion{H}{i} absorption-line gas. Figures \ref{3c49mod} and \ref{3c268mod} show the clean maps, the Gaussian components used to model them, and the location of the \ion{H}{i} absorption.  The coordinates (0,0) were driven by self-calibration to coincide with the brightest pixel in the image, rather than with the location of the brightest Gaussian component. Tables \ref{measurements1} and \ref{measurements2} list the details of all the Gaussian source components and their fitted absorption line properties.\\

\subsection{The Milliarcsecond Scale Continuum Structure}
\label{subsec:continuum}

These radio sources have been been previously imaged with high angular resolution \citep[e.g.,][]{Ludke98, Sanghera95, Neff95, Breugel92, Nan91, Akujor91}.  Our small array of telescopes did not allow us to perform a high quality calibration but we obtained continuum maps and structures consistent with those previously published.\\

\subsubsection{\object{3C~49}}

\object{3C~49} is a CSS galaxy \citep{Spinrad85}  at a redshift z=0.6207 with radio components extended mostly east-west over a total angular size  of 1 arcsec or $\simeq$6.8 kpc\footnote{Ho = 71 km s\mone Mpc\mone, $\Omega_M=0.27$ and $\Omega_\Lambda=0.73$ }. \\

We detect the three components (labeled East, Center-West (CW) and West) which have been seen previously \citep[i.e.][]{Nan91, Breugel92, Neff95, Sanghera95, Ludke98}. Multifrequency observations have shown that CW contains the flat spectrum core \citep{Sanghera95,Ludke98}. We also find an additional component labeled Center-East (CE) whose reality is uncertain. The CE component is faint  but its inclusion in the clean components seemed to improve the final image. West is the brightest component in the map (integrated flux density = 7.48 $\pm$ 0.06 Jy), which is roughly 66, 33 and 4.5 times more than the integrated flux density of  CE, CW and East respectively .\\ 


\subsubsection{\object{3C~268.3}}

\object{3C~268.3} is a  CSS galaxy \citep{Spinrad85, Gelderman94} at a redshift z=0.37116, with two main radio components oriented roughly north-south with a total angular size of 1.5 arcsec or $\simeq$7.6 kpc \citep[i.e.,][]{Nan91, Breugel92, Neff95, Sanghera95, Ludke98}. Our observations are consistent with the previous results.  \citet{Ludke98} have found a faint, possible core at 5 GHz, about 1.5 kpc south of the northern lobe.\\ 

The northern component has an integrated flux density of 2.25$\pm$0.01 Jy, which is roughly 22 times more than the southern  component. The northern component also shows a more complex structure: 6 Gaussian components were required for it, as opposed to a single one for the southern  feature.\\

As for the components of \object{3C~49}, an offset between the brightest Gaussian component and the image peak brightness of \object{3C~268.3} is observed. In this case, the offset is even bigger, which is consistent with the North component being reproduced by more, brighter and bigger Gaussian components than W in \object{3C~49}.\\


\subsection{The  \ion{H}{i} Absorption}
\label{sec:absorption}

We have detected 21 cm \ion{H}{i} absorption against the western lobe of \object{3C~49} and the northern lobe of \object{3C~268.3}. Figures \ref{spec1}, \ref{spec3} and \ref{tau1}, \ref{tau3} show the spectra and optical depth of these regions in \object{3C~49} and \object{3C~268.3}.\\

The measured peak depth $\Delta$S of the absorption line depends on both the optical depth $\tau$ and the covering factor of the hydrogen c$_f$ \citep[e.g.][]{Wolfe75},

\begin{center} 
 c$_f = \frac{\Delta\mathrm{S}}{\mathrm{S} (1 - e^{-\tau})}$  \hfil or \hfil   $\tau = - \mathrm{ln} \left( 1 - \frac{\Delta\mathrm{S}}{\mathrm{Sc}_f}  \right)$ 
\end{center}

where S is the continuum flux density. For a uniform source there is a minimum covering factor required by the observed ratio of line depth to continuum flux density c$_f >  \Delta$S / S \citep[e.g.][]{O'Dea94}, and for a complete covering of the source c$_f=1$. \\

The column density, $N_{\ion{H}{i}}$, is given by:

\begin{center}
 $N_\mathrm{HI}$ = 0.18 x 10$^{21}$ ($T_\mathrm{S}$/100K) $\int \tau_v ~ dv$ cm$^{-2} $
\end{center}

\citep[e.g.,][]{Dopita03, O'Dea94}, where $T_s$ is the spin temperature, $\tau$ the optical depth, and {\it v} is the velocity. For a Gaussian profile:

\begin{center}
$N_{\ion{H}{i}} \simeq $ 1.94 x 10$^{20}$ ($T_\mathrm{S}$/100K) $\tau_0 \Delta V $cm$^{-2} $
\end{center}

where $\tau_0$ is the peak optical depth in the line and $\Delta V$ is the FWHM of the Gaussian line profile. A spin temperature T$_\mathrm{s}$ = 100 K is applicable under typical ISM circumstances, although close to an AGN it could be higher \citep[e.g.][]{Morganti01}.\\

\subsubsection{\object{3C~49}}

\object{3C~49} shows \ion{H}{i} absorption in the center of the Western component with a peak depth of $\sim$4$\%$ of the continuum level. The reduced $\chi^2$ indicated that two Gaussian lines were needed to fit the absorption with peak depths of $\sim$1.5$\%$ and $\sim$2.1$\%$. The column densities are 1.5 and 0.8 $ \times 10^{20}$ ($T_\mathrm{S}$/100K) cm$^{-2}$. The absorption seems to be blueshifted in the host galaxy rest frame.\\ 

The covering factor, c$_f$ for the absorption towards W in \object{3C~49} is $1 > c_f > 0.04$. If we assume the two absorption lines to be produced by spherical clouds of similar radius, equal to the measured FWHM of the component (24 mas = 163 pc at the source distance), then the density of these clouds would be 220/c$_f$ (T$_\mathrm{s}$/100) cm$^{-3}$ and 160/c$_f$  (T$_\mathrm{s}$/100) cm$^{-3}$. \\


\subsubsection{\object{3C~268.3}}

\object{3C~268.3} shows an absorption profile in the center of the northern lobe. $\chi^2$ fitting shows that an adequate representation for the detection is just one Gaussian line redshifted in the host galaxy rest frame; the column depth is 3.2x10$^{20}$ ($T_\mathrm{S}$/100K) cm$^{-2}$. Inspection of Table \ref{measurements2} shows that the limits on the other components are too high to rule out the presence of \ion{H}{i} at similar levels in the southern  component, which is much fainter than the northern component. The most intriguing 2$\sigma$ detection limit is 3.0\% in component 4(N), which, compared to the detection in component 3(N), suggests that, possibly, the extent of the absorption towards the northern lobe is limited to $72\times39$ mas ($367\times199$ pc at the source distance).\\

The covering factor, c$_f$, for \object{3C~2683} is $1 > c_f > 0.025$. If we assume the absorption line to be produced by a spherical cloud of a radius equal to the average of the measured FWHM of the component (55.5 mas = 283 pc), then the density of the cloud would be 360/c$_f$ (T$_\mathrm{S}$/100) cm$^{-3}$.\\

\section{Discussion}
\label{sec:discussion}

We have detected \ion{H}{i} absorption towards the western radio lobe of \object{3C~49} and the northern lobe of \object{3C~268.3}. There are several possible hypotheses for the nature of the absorbing gas; e.g., (1) unrelated foreground clouds (e.g., in the ISM of the host galaxy) (2) an organized structure in the host galaxy such as a disk \citep[as discussed by][]{Pihlstrom03}, and (3) clouds which are interacting with the radio jet. We argue here that our results are consistent with the \ion{H}{i} being produced in clouds which are in the environment of the radio source.  \\

In \object{3C~49},  the west lobe is  $\sim 4.5$ times brighter than the East lobe and $\sim 3$ times closer to the core. Both radio lobes are unpolarized \citep{Ludke98}. The emission line gas imaged by HST is very faint but seems to be  asymmetric and is brighter near the west lobe \citep{Vries97, Vries99, Axon00}. \\


In \object{3C~268.3}, the northern radio lobe is roughly 22 times brighter than the southern lobe and $\sim 3$ times closer to the core. In addition, \citet{Ludke98} show that the radio polarization is asymmetric and that the northern lobe is much more strongly depolarized than the southern lobe. The emission line gas imaged by HST is also asymmetrically distributed and is much brighter near the northern lobe, while the southern lobe seems to extend beyond the line emission \citep{Vries97, Vries99, Axon00}.\\


These correlated asymmetries in the radio and optical line emission are similar to those seen by \citet{McCarthy91} in samples of powerful high redshift radio galaxies. We suggest that these data are consistent with a picture in which one side of the radio source is strongly interacting with dense clouds of gas. The interaction causes the lobe to propagate more slowly, resulting in a smaller separation distance from the core \citep[e.g.,][]{Young91, Carvalho98, Jeyakumar05}. The higher radio luminosity could be due to smaller adiabatic expansion losses or to increased energy production efficiency due to compression and shocks \citep[e.g.,][]{Jeyakumar05}. The clouds will also act as a Faraday screen, causing the observed depolarization as seen in \object{3C~268.3}. The association of the optical emission line clouds and the \ion{H}{i} absorption suggests that the \ion{H}{i} absorption is produced in the atomic cores of the clouds which are seen in the emission line images \citep[as may be the case in PKS2322-123,][]{O'Dea94}. However, we caution that the data do not exclude the existence of \ion{H}{i} on both sides of \object{3C~49} and \object{3C~268.3}. An alternate scenario would be that the brighter lobe is beamed towards us and we are seeing \ion{H}{i} absorption produced in clouds which are swept up by the radio lobes. However, if the brightness asymmetry is due to beaming we would expect the brighter lobe to be further from the core than the weaker lobe (due to the difference in light travel time), which is the opposite of what is observed. \\

A study of the emission line nebulae in three CSS sources -- \object{3C~67}, \object{3C~277.1} and \object{3C~303.1} -- using HST long slit spectroscopy has shown that the kinematics of the gas are consistent with the clouds having been accelerated to velocities of several hundred km/s by shocks induced by the expanding radio lobes \citep{O'Dea02}. The ionization diagnostics of the gas in CSS sources are also consistent with a contribution from shock-ionized gas \citep{Labiano05, Morganti97, Gelderman94}. Thus, the observed blue and redshifts of the \ion{H}{i} clouds of several hundred km/s in \object{3C~49} and \object{3C~268.3} could be attributed to bow shock induced velocities. It seems likely  that both positive and negative velocities can result, depending on whether the \ion{H}{i} absorption occurs in gas pushed towards the observer or in gas being entrained away from the observer.\\

\section{Summary}
\label{sec:summary}

We present European VLBI Network spectral line observations  in the UHF band of the  redshifted 21 cm \ion{H}{i} line in two compact steep spectrum radio galaxies. We have detected  \ion{H}{i} absorption towards the western radio lobe of \object{3C~49} and the northern lobe of \object{3C~268.3}. The radio  lobes with \ion{H}{i} absorption (1) are brighter and closer to the core than the opposite lobes; (2) are more depolarized (in \object{3C~268.3}); and (3) are preferentially associated with optical emission line gas. The association between the \ion{H}{i} absorption and the emission line gas, supports the hypothesis that the \ion{H}{i} absorption is produced in the atomic cores of the emission line clouds, but we cannot rule out the existence of \ion{H}{i} elsewhere. We suggest that the asymmetries in the radio and optical emission are due to interaction of the radio source with an asymmetric distribution of dense clouds in their environment. Our results are consistent with a picture in which CSS sources interact with clouds of dense gas as they propagate through their host galaxy.\\

\begin{acknowledgements}

This research has made use of the NASA/IPAC Extragalactic Database (NED) which is operated by the Jet Propulsion Laboratory, California Institute of Technology, under contract with the National Aeronautics and Space Administration. This research has made use of NASA's Astrophysics Data System. The European VLBI Network is a joint facility of European, Chinese, South African and other radio astronomy institutes funded by their national research councils. \\ 

\end{acknowledgements}

\bibliographystyle{aa}
\bibliography{3856Refs}



\end{document}